\documentclass[10pt, conference, letterpaper]{IEEEtran}
\IEEEoverridecommandlockouts
\usepackage{cite}
\usepackage{amsmath,amssymb,amsfonts}
\usepackage{algorithmic}
\usepackage{graphicx}
\usepackage{textcomp}
\usepackage{xcolor}
\usepackage{subcaption}
\usepackage{gensymb}
\usepackage{hyperref}
\usepackage{multirow}
\usepackage[export]{adjustbox}
\usepackage[right=0.631in, left=0.631in, top= 0.751in, bottom= 1.1in]{geometry}

\def\BibTeX{{\rm B\kern-.05em{\sc i\kern-.025em b}\kern-.08em
    T\kern-.1667em\lower.7ex\hbox{E}\kern-.125emX}}
\begin{document}

\title{Localization with Noisy Android Raw GNSS Measurements\\
\thanks{This work was supported by Advanced Manufacturing and Engineering (AME) Industry Alignment Fund—Pre-Positioning (IAF-PP) (Grant No. A19D6a0053) funded by Agency for Science, Technology and Research (A*STAR) in Research, Innovation and Enterprise (RIE) 2020, Singapore.}
}
\author{
\IEEEauthorblockN{Xu Weng}
\IEEEauthorblockA{\textit{School of Electrical and Electronic Engineering} \\
\textit{Nanyang Technological University}\\
Singapore \\
xu009@e.ntu.edu.sg}
\and
\IEEEauthorblockN{KV Ling}
\IEEEauthorblockA{\textit{School of Electrical and Electronic Engineering} \\
\textit{Nanyang Technological University}\\
Singapore \\
ekvling@ntu.edu.sg}
}

\maketitle

\begin{abstract}
Android raw Global Navigation Satellite System (GNSS) measurements are expected to bring smartphones power to take on demanding localization tasks that are traditionally performed by specialized GNSS receivers. The hardware constraints, however, make Android raw GNSS measurements much noisier than geodetic-quality ones. This study elucidates the principles of localization using Android raw GNSS measurements and leverages Moving Horizon Estimation (MHE), Extended Kalman Filter (EKF), and Rauch-Tung-Striebel (RTS) smoother for noise suppression. Experimental results show that the RTS smoother achieves the best positioning performance, with horizontal positioning errors significantly reduced by 76.4\% and 46.5\% in static and dynamic scenarios compared with the baseline weighted least squares (WLS) method. Our codes are available at https://github.com/ailocar/androidGnss.
\end{abstract}


\section{Introduction}
Localization is an essential technology that underpins the interconnected world. Global Navigation Satellite Systems (GNSS) provide location information that drives industries ranging from defense and agriculture to geomatics and transportation. However, high-performance positioning services are still far away from our daily lives because specialized GNSS equipment is bulky and expensive. Unlike dedicated GNSS devices, ubiquitous portable smartphones integrate affordable GNSS chips and antennas, providing huge potential for everyday location services. Especially since the release of Android raw GNSS measurements, smartphones have been expected to enable various exciting localization-based applications, such as vehicle navigation, intelligent management of city assets, outdoor augmented reality, and mobile health monitoring. Nevertheless, it is difficult to keep such promises due to the large noise present in these measurements collected by current mass-market Android devices.

Researchers around the world have conducted systematic assessments of Android raw GNSS measurements. It is reported that the average carrier-to-noise power density ($C/N_0$) of smartphones' Global Positioning System (GPS) L1 observations is about ten dB$\cdot$Hz lower than the representative value for geodetic antennas and receivers \cite{zhang2018quality}. Consequently, the pseudorange noise of phones is about an order of magnitude larger than geodetic-quality measurements \cite{li2019characteristics}. Therefore, Android raw GNSS measurements must be denoised to meet the essential requirement of potential mobile localization applications.

\begin{figure}[!t]
    \centering
    \includegraphics[width=0.35\textwidth]{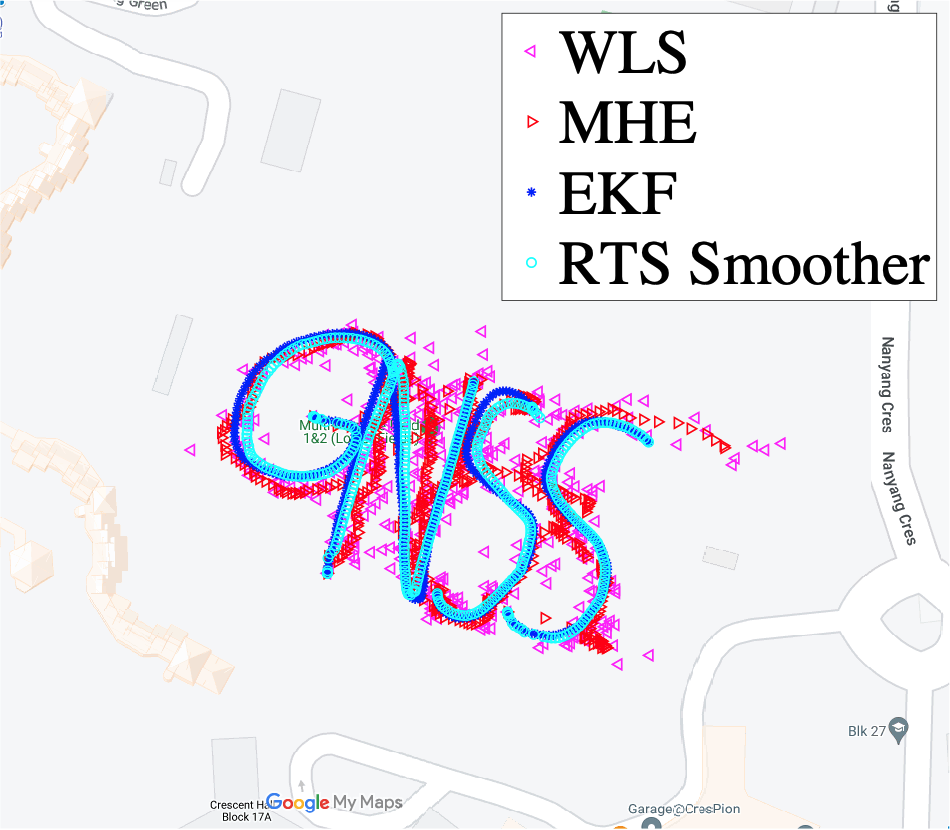}
    \caption{Localization results using the Weighted Least Squares (WLS) method, Moving Horizon Estimation (MHE), Extended Kalman Filter (EKF), and Rauch-Tung-Striebel (RTS) smoother with Android raw GNSS measurements collected by Xiaomi Redmi K60}
    \label{fig: toyEx}
\end{figure}

The Extended Kalman Filter (EKF) is widely applied to denoise pseudorange measurements for GNSS receivers \cite{spilker1996global,kaplan2017understanding}. For example, an EKF-based GNSS positioning approach using Android data for maritime applications has been demonstrated in \cite{innac2021kalman}, but it fails to give a complete description of how to construct the process model and measurement equations with pseudoranges and pseudorange rates. A Least Squares (LS) localization engine followed by a Kalman filter is introduced in \cite{jiang2022open}, but its connections to Android raw GNSS measurements are not explained in detail. Another work has proposed an EKF with inequality constraints, including vertical velocity, direction, and distance constraints, to process Android GNSS data \cite{peng2023improving}. Its experimental results show obvious improvement in localization performance credited to these constraints. However, like the aforementioned studies, there are no clear specifications of its implementation over Android raw GNSS measurements.

Moving Horizon Estimation (MHE) also shows potential for state estimation from noisy measurements \cite{ling1999receding}. Some sophisticated tracking loops have been designed to enhance the performance of GNSS receivers in challenging scenarios involving high dynamics, signal fading, strong ionosphere scintillation, and so on \cite{wang2015robust,qin2019state}. In the measurement domain, MHE has been applied to vehicular localization using GNSS measurements aided by map boundaries \cite{brembeck2019nonlinear}. Besides, in the area of high-accuracy positioning, a state‐space‐varied MHE algorithm is proposed to solve inconstant states present in precise point positioning (PPP) algorithms for low-cost receivers and Android smartphones in harsh environments \cite{liu2023state}. The prior research has focused on the design of advanced algorithms but ignores the implementation details. And the MHE-based pseudorange positioning with Android raw GNSS measurements has yet to be fully covered.

In addition to these filtering methods, the Rauch-Tung-Striebel (RTS) smoother, a post-processing algorithm, also delivers exceptional noise mitigation ability \cite{rauch1965maximum}. RTS smoother has been applied to integrated navigation systems \cite{chiang2012line, siemuri2022application}. Nevertheless, when it comes to the RTS smoother-based localization using Android raw GNSS measurements, only some codes are available on the internet \cite{kaggleNotebook}. A systematic and technical study about it is still scarce in the GNSS community.

Previous research has rarely provided a detailed and systematic introduction to pseudorange-based positioning using noisy Android raw GNSS measurements, especially engineering details about adapting the well-established filtering or smoothing theories to Android data, including measurement acquisition, noise modeling, and exception handling. Unlike studies above that concentrate on the design of filtering or smoothing algorithms, our work thoroughly explains the adaptation of various filtering or smoothing algorithms to noisy Android raw GNSS measurements from an engineering perspective. Our main contributions are listed below:
\begin{itemize}
    \item We detail how to calculate position, velocity, and time (PVT) using Android raw GNSS measurements. 
    \item We employ MHE, EKF, and RTS smoother to filter or smooth noisy Android raw GNSS measurements to improve localization performance. We design finite state machines for these algorithms to address Android data discontinuity. To the best of our knowledge, our work is the first to systematically go into detail about how to filter or smooth Android raw GNSS measurements. 
    \item We evaluate these algorithms using static data we collected and dynamic data from Google.
\end{itemize}

\section{Determining Position, Velocity, and Time using Android Raw GNSS Measurements}
This section details the pseudorange-based position, velocity, and time (PVT) determination using the Weighted Least Squares (WLS) algorithm with Android raw GNSS measurements.
\subsection{Calculating Pseudorange Measurements with Android Raw GNSS Measurements}
Android phones record the moments related to the propagation of GNSS signals. The time when GNSS signals are transmitted $t_{Tx}$ is logged as \emph{ReceivedSvTimeNanos} in nanoseconds in their respective GNSS time reference systems, i.e.,
\begin{equation*}\label{eq:tTx}
t_{Tx} = ReceivedSvTimeNanos - \Delta t_{constellation}
\end{equation*}
where $\Delta t_{constellation}$ represents the time difference between the current constellation time and the GPS reference time. The time when Android smartphones receive GNSS signals is calculated as follows: 
\begin{IEEEeqnarray*}{rCl}\label{eq:tRx}
t_{Rx}&=&TimeNanos + TimeOffsetNanos \nonumber
\\
&&- \left(FullBiasNanos(1)+BiasNanos(1)\right) \nonumber
\\
&&-weekNumberNanos
\end{IEEEeqnarray*}
where \emph{TimeNanos} is the internal hardware clock of Android phones in nanoseconds. \emph{TimeOffsetNanos} is the offset between $TimeNanos$ and the actual measurement time. \emph{FullBiasNanos} estimates the difference between smartphone clocks and GPS time in full nanoseconds, while \emph{BiasNanos} is the sub-nanosecond section. Here, we use the initial values of \emph{FullBiasNanos} and \emph{BiasNanos} to include the hardware clock drift into $t_{Rx}$. \emph{weekNumberNanos} represents the total full weeks in nanoseconds since midnight on January 5-6, 1980, and is calculated as follows:
\begin{IEEEeqnarray*}{rCl}
weekNumberNanos&=&\lfloor\frac{-FullBiasNanos}{NumberNanoSecondsWeek}\rfloor \nonumber
\\
&&\times NumberNanoSecondsWeek
\end{IEEEeqnarray*}
where \emph{NumberNanoSecondsWeek} represents the total nanoseconds in a week. Because \emph{FullBiasNanos} is a negative value, there is a minus sign before it.

After obtaining the moments when GNSS signals are transmitted and received, the pseudorange measurements $\rho$ can be calculated using the speed of light $c$ as follows: 
\begin{equation*}\label{eq:prAndroid}
\rho = \left(t_{Rx} - t_{Tx} \right)c. 
\end{equation*}  

Note that $t_{Tx}$ and $t_{Rx}$ are in the format of time of week (TOW), so the week rollover should be considered when we calculate the difference between them. 

Android phones directly provide the pseudorange rate measurement \emph{PseudorangeRateMetersPerSecond} and its 1-$\sigma$ uncertainty \emph{PseudorangeRateUncertaintyMetersPerSecond}.

\subsection{Pseudorange and Pseudorange Rate Models}
After removing the satellite clock offset and atmosphere delays, which can be modeled and computed using broadcast ephemeris, we are able to write the corrected pseudorange $\rho_{c_k}^{(n)}$  from the $n^{th}$ satellite to a phone at the $k^{th}$ epoch as
\begin{equation}\label{eq:cpr}
\rho_{c_k}^{(n)}=r_k^{(n)}+\delta t_{u_k}+\varepsilon_k^{(n)}
\end{equation}
where $r_k^{(n)}$ denotes the geometry range from the $n^{th}$ satellite to the phone. $\delta t_{u_k}$ represents the clock offset of the phone relative to the GNSS reference time. We wrap up the multipath delay, hardware delay, pseudorange noise, modeling residuals of atmosphere delays, and other potential errors in one item $\varepsilon_k^{(n)}$ called pseudorange errors. We can compute the corrected pseudorange rate measurement $\Dot{\rho}_{c_k}^{(n)}$ by removing the satellite clock drift that is the derivative of the satellite clock bias, which is shown below:
\begin{equation}\label{eq:cprr}
\Dot{\rho}_{c_k}^{(n)}=\Dot{r}_k^{(n)}+\delta f_{u_k} + \Dot{\varepsilon}_{k}^{(n)}
\end{equation}
where $\dot{r}_k^{(n)}$ represents the geometry range rate from the $n^{th}$ satellite to the user. $\delta f_{u_k}$ represents the clock drift of the user device. $\dot{\varepsilon}_{k}^{(n)}$ denotes the variation of pseudorange errors.

\subsection{WLS-based PVT Solution}
At the $k^{th}$ epoch, an Android phone's position $\left(x_k, y_k, z_k\right)$, velocity $\left(v_{x_k}, v_{y_k}, v_{z_k}\right)$, clock offset $\delta t_{u_k}$, and clock drift $\delta f_{u_k}$ are unknowns to be estimated. We can estimate all of them simultaneously using the WLS algorithm. Let $\mathbf{X}_k=\left[x_k,v_{x_k},y_k,v_{y_k},z_k,v_{z_k}, \delta t_{u_k},\delta f_{u_k}\right]^T$. The location and velocity of the $n^{th}$ satellite are denoted by $\mathbf{x}_k^{(n)}=[x_k^{(n)}, y_k^{(n)}, z_k^{(n)}]^T$ and $\mathbf{v}_k^{(n)} = [v_{x_k}^{(n)}, v_{y_k}^{(n)}, v_{z_k}^{(n)}]^T$ respectively, which can be derived from ephemeris data. If the Android phone receives signals transmitted by $M$ satellites, $2M$ measurements like \eqref{eq:cpr} and \eqref{eq:cprr} can be collected. And if we know an approximation of the phone's state, i.e., $\tilde{\mathbf{X}}_k=[\tilde{x}_k,\tilde{v}_{x_k}, \tilde{y}_k,\tilde{v}_{y_k},\tilde{z}_k,\tilde{v}_{z_k},\delta \tilde{t}_{u_k},\delta \tilde{f}_{u_k}]^T$, the PVT solution can be found by solving the following linear equation system \cite{kaplan2017understanding}:
\begin{equation}\label{eq:PosEq}   
\mathbf{G}_k\left(\mathbf{X}_k-\tilde{\mathbf{X}}_k\right)=\mathbf{b}_k  
\end{equation}
where
\begin{gather}
\mathbf{G}_k=\begin{bmatrix}\tilde{a}_{x_k}^{(1)}&0&\tilde{a}_{y_k}^{(1)}&0&\tilde{a}_{z_k}^{(1)}&0&1&0
\\
                0&\tilde{a}_{x_k}^{(1)}&0&\tilde{a}_{y_k}^{(1)}&0&\tilde{a}_{z_k}^{(1)}&0&1
                \\
                \vdots&\vdots&\vdots&\vdots&\vdots&\vdots&\vdots&\vdots
                \\
                \tilde{a}_{x_k}^{(M)}&0&\tilde{a}_{y_k}^{(M)}&0&\tilde{a}_{z_k}^{(M)}&0&1&0
                \\
                0&\tilde{a}_{x_k}^{(M)}&0&\tilde{a}_{y_k}^{(M)}&0&\tilde{a}_{z_k}^{(M)}&0&1
                \end{bmatrix}\nonumber
\\
\tilde{a}_{x_k}^{(n)}=\frac{\tilde{x}_k-x_k^{(n)}}{\tilde{r}_k^{(n)}},\tilde{a}_{y_k}^{(n)}=\frac{\tilde{y}_k-y_k^{(n)}}{\tilde{r}_k^{(n)}},\tilde{a}_{z_k}^{(n)}=\frac{\tilde{z}_k-z_k^{(n)}}{\tilde{r}_k^{(n)}} \nonumber
\end{gather}
\begin{gather}
\tilde{r}_k^{(n)} = \sqrt{\left(\tilde{x}_k-x_k^{(n)}\right)^2+\left(\tilde{y}_k-y_k^{(n)}\right)^2+\left(\tilde{z}_k-z_k^{(n)}\right)^2}\nonumber
\\
\mathbf{b}_k=[\Delta \rho_{c_k}^{(1)},\Delta\Dot{\rho}_{c_k}^{(1)},\Delta \rho_{c_k}^{(2)},\Delta\Dot{\rho}_{c_k}^{(2)},\cdots,\Delta \rho_{c_k}^{(M)},\Delta\Dot{\rho}_{c_k}^{(M)}]^T \nonumber
\\
\Delta \rho_{c_k}^{(n)}=\rho_{c_k}^{(n)}-\tilde{r}_k^{(n)}-\delta \tilde{t}_{u_k}\nonumber
\\
\Delta\Dot{\rho}_{c_k}^{(n)}=\Dot{\rho}_{c_k}^{(n)} -\left(\Tilde{\mathbf{v}}_k-\mathbf{v}_k^{(n)}\right)\cdot\tilde{\mathbf{g}}_k^{(n)}-\delta \Tilde{f}_{u_k} \nonumber
\\
\Tilde{\mathbf{v}}_k = \left[\tilde{v}_{x_k},\tilde{v}_{y_k},\tilde{v}_{z_k}\right]^T\nonumber
\\
\tilde{\mathbf{g}}_k^{(n)}=[\tilde{a}_{x_k}^{(n)},\tilde{a}_{y_k}^{(n)}, \tilde{a}_{z_k}^{(n)}]^T. \nonumber
\end{gather}

To balance the impact of noise on the estimation precision, 1-$\sigma$ uncertainties of pseudorange and pseudorange rate measurements can be used to weight \eqref{eq:PosEq} as follows: 
\begin{equation}\label{eq:WPosEq}   
\mathbf{W}_k\mathbf{G}_k\left(\mathbf{X}_k-\tilde{\mathbf{X}}_k\right)=\mathbf{W}_k\mathbf{b}_k
\end{equation}
where $\mathbf{W}_k$ is a diagonal weight matrix with the reciprocals of 1-$\sigma$ uncertainties of pseudorange and pseudorange rate measurements of different satellites as its main diagonal. The 1-$\sigma$ uncertainty of $t_{Tx}$ given by \emph{ReceivedSvTimeUncertaintyNanos} can represent the 1-$\sigma$ pseudorange measurement uncertainty. The 1-$\sigma$ pseudorange rate uncertainty is given by \emph{PseudorangeRateUncertaintyMetersPerSecond}. Then, the WLS solution to \eqref{eq:WPosEq} can be computed as \cite{kaplan2017understanding}
\begin{IEEEeqnarray}{rCl}\label{eq:xk}
    \mathbf{X}_k&=&\tilde{\mathbf{X}}_k+\Delta \mathbf{X}_k \nonumber
    \\   &=&\tilde{\mathbf{X}}_k+\left(\mathbf{W}_k\mathbf{G}_k\right)^+\mathbf{W}_k\mathbf{b}_k
\end{IEEEeqnarray}
where $\Delta\mathbf{X}_k=\left(\mathbf{W}_k\mathbf{G}_k\right)^+\mathbf{W}_k\mathbf{b}_k$ is the displacement from the approximate user state to the actual one. The approximate user state $\tilde{\mathbf{X}}_k$ will be updated with the result of \eqref{eq:xk}, and the computation in \eqref{eq:xk} will be iterated until the accuracy requirement is satisfied. Note that the approximation of state $\tilde{\mathbf{X}}_k$ can be initialized as all zeros or set as the phone's state at the last epoch. 

\section{Modeling Android Phones}\label{sec:modeling}
\subsection{Process Models of Android Phones}
\begin{figure}[!t]
    \centering
    \includegraphics[scale=0.18]{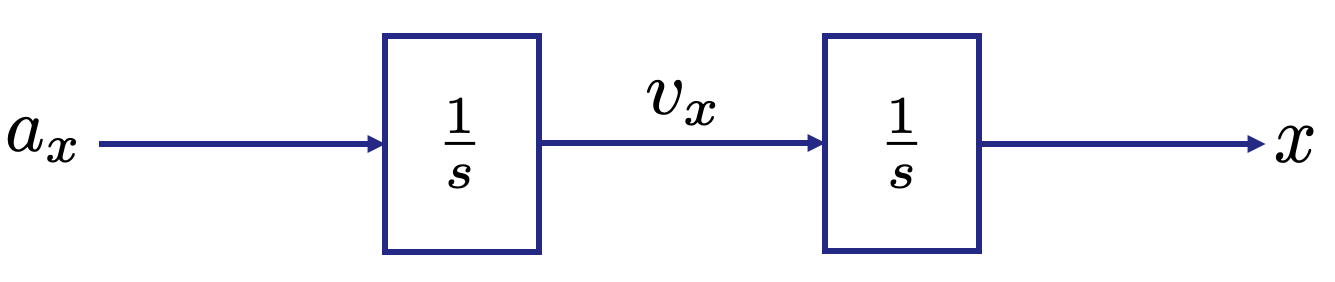}
    \caption{Dynamic model for Android phones}
    \label{fig:dynamic_model}
\end{figure}

A suitable process model can connect the states of Android phones at various time steps. Considering the low dynamics of Android phones, their states can be described by a two-state model. For example, the dynamic modal of the phone's location and velocity on the $x$ axis is shown in Fig. \ref{fig:dynamic_model}, where $a_x$ is the acceleration on the $x$ axis. Accordingly, the discrete dynamic modal can be written as
\begin{equation}\label{eq:discretDynamicEq}
        \mathbf{X}_{x_k} =\mathbf{A}_{t_k,t_{k-1}}^{(x)}\mathbf{X}_{x_{k-1}}+\mathbf{W}_{x_{k-1}}
\end{equation}
where
\begin{gather}
    \mathbf{X}_{x_k} = \left[x_k,v_x\right]^T \nonumber
    \\
    \mathbf{A}_{t_k,t_{k-1}}^{(x)}=\begin{bmatrix}1&T_{s_k}\\0&1
    \end{bmatrix}\label{eq:Ax} 
    \\ 
    \mathbf{Q}_{x_{k-1}}={\rm E}\left(\mathbf{W}_{x_{k-1}}\mathbf{W}_{x_{k-1}}^T\right)= \begin{bmatrix}
        \frac{1}{3}S_{v_x}T_{s_k}^3 & \frac{1}{2}S_{v_x}T_{s_k}^2\\
        \frac{1}{2}S_{v_x}T_{s_k}^2 & S_{v_x}T_{s_k}
    \end{bmatrix} \nonumber
    \\
    S_{v_x}={\rm E}(a_{x}^2). \nonumber
\end{gather}
$T_{s_k}$ is the sampling period of the system at the $k^{th}$ epoch. $S_{v_x}$ can be estimated as follows:
\begin{equation}
    S_{v_x}={\rm E}(a_{x}^2) \approx a_{x}^2 \approx \left(\frac{\hat{v}_{x_{k-1}}-\hat{v}_{x_{k-2}}}{T_{s_{k-1}}}\right)^2 \label{eq:SxCompute}
\end{equation}
where $\hat{v}_{x_{k-1}}$ and $\hat{v}_{x_{k-2}}$ is the estimated velocities at previous epochs. The same dynamic models can be established for motions on the $y$ and $z$ axes. 

\begin{figure}[!t]
    \centering
    \includegraphics[scale=0.18]{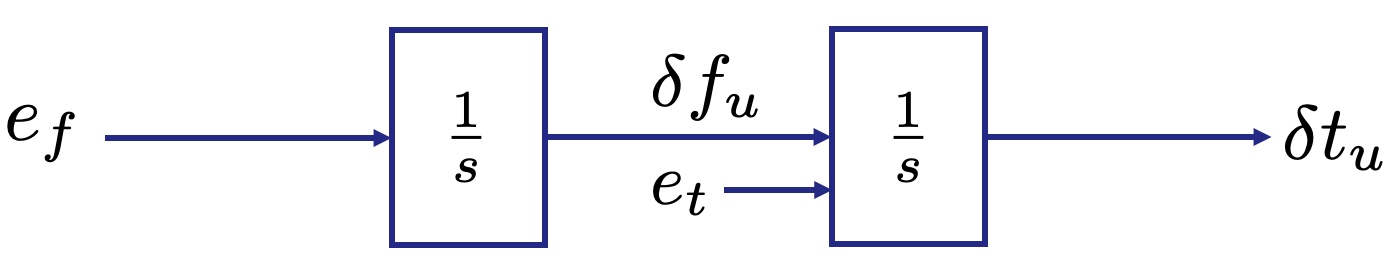}
    \caption{Dynamic model for hardware clock of Android phones}
    \label{fig:clock_model}
\end{figure}

The dynamic model for the hardware clock of Android phones can be represented by Fig. \ref{fig:clock_model}, where $e_t$ and $e_f$ are the clock offset noise and the clock drift noise. Accordingly, the discrete dynamic model for the clock states of Android phones can be written as
\begin{equation}\label{eq:discretClockEq}
        \mathbf{X}_{t_k} =\mathbf{A}_{t_k,t_{k-1}}^{(t)}\mathbf{X}_{t_{k-1}}+\mathbf{W}_{t_{k-1}}
\end{equation}
where
\begin{gather}
    \mathbf{X}_t = \left[\delta t_u,\delta f_u\right]^T \nonumber
    \\
    \mathbf{A}_{t_k,t_{k-1}}^{(t)} = \begin{bmatrix}1&T_{s_k}\\0&1
    \end{bmatrix}\label{eq:At} 
    \\
    \mathbf{Q}_{t_{k-1}}={\rm E}\left(\mathbf{W}_{t_{k-1}}\mathbf{W}_{t_{k-1}}^T\right)= \begin{bmatrix}
        S_tT_{s_k}+\frac{1}{3}S_fT_{s_k}^3 & \frac{1}{2}S_fT_{s_k}^2\\
        \frac{1}{2}S_fT_{s_k}^2 & S_fT_{s_k}
    \end{bmatrix}\label{eq:Qt} \nonumber
    \\
    S_t={\rm E}(e_t^2),\ S_f={\rm E}(e_f^2). \nonumber
\end{gather}

According the clock model illustrated by Fig. \ref{fig:clock_model}, $S_t$ and $S_f$ can be simply estimated as follows:
\begin{gather}
    S_{t}={\rm E}(e_{t}^2) \approx e_{t}^2 \approx \left(\frac{\delta \hat{t}_{u_{k-1}}-\delta \hat{t}_{u_{k-2}}}{T_{s_{k-1}}}-\delta \hat{f}_{u_{k-1}}\right)^2 \label{eq:StCompute} 
    \\
    S_{f} = {\rm E}(e_{f}^2) \approx e_{f}^2 \approx \left(\frac{\delta \hat{f}_{u_{k-1}}-\delta \hat{f}_{u_{k-2}}}{T_{s_{k-1}}}\right)^2 \label{eq:SfCompute} 
\end{gather}
where $\delta\hat{t}_{u_{k-1}}$, $\delta\hat{t}_{u_{k-2}}$, $\delta\hat{f}_{u_{k-1}}$, and $\delta\hat{f}_{u_{k-1}}$ are previously estimated clock bias and clock drift. 

According to \eqref{eq:discretDynamicEq} and \eqref{eq:discretClockEq}, the joint process model of an Android phone can be written as
\begin{equation}\label{eq:stateUpdate}
    \mathbf{X}_{k}=\mathbf{A}_{k,k-1}\mathbf{X}_{k-1}+\mathbf{W}_{k-1}
\end{equation}
where
\begin{equation}
    \mathbf{A}_{k,k-1} = \begin{bmatrix}\mathbf{A}_{t_k,t_{k-1}}^{(x)}&0&0&0\\ 0&\mathbf{A}_{t_k,t_{k-1}}^{(y)}&0&0\\ 0&0&\mathbf{A}_{t_k,t_{k-1}}^{(z)}&0\\ 0&0&0&\mathbf{A}_{t_k,t_{k-1}}^{(t)} \end{bmatrix} \label{eq:stateTransMatrix} 
\end{equation}
\begin{IEEEeqnarray*}{rCl}
    \mathbf{Q}_{k-1} &=& {\rm E}\left(\mathbf{W}_{k-1}{\mathbf{W}_{k-1}}^T\right) \nonumber
    \\
    &=&\begin{bmatrix}\mathbf{Q}_{x_{k-1}}&0&0&0\\ 0&\mathbf{Q}_{y_{k-1}}&0&0\\ 0&0&\mathbf{Q}_{z_{k-1}}&0\\ 0&0&0&\mathbf{Q}_{t_{k-1}} \end{bmatrix}. \nonumber
\end{IEEEeqnarray*}

\subsection{Measurement Models of Android Phones}
For Android phones in low dynamics, according to \eqref{eq:cpr}, \eqref{eq:cprr}, and \eqref{eq:PosEq}, we can get the joint pseudorange and pseudorange rate measurement equations as follows:
\begin{equation}\label{eq:measurementEqn}
    \mathbf{b}_k=\mathbf{C}_k(\mathbf{X}_{k}-\Tilde{\mathbf{X}}_{k})+\mathbf{E}_k
\end{equation}
where
\begin{gather}
    \mathbf{C}_k = \mathbf{G}_k \nonumber
    \\
     \mathbf{E}_k = [\varepsilon_k^{(1)},\Dot{\varepsilon}_{k}^{(1)},\varepsilon_k^{(2)},\Dot{\varepsilon}_{k}^{(2)},\cdots,\varepsilon_k^{(M)},\Dot{\varepsilon}_{k}^{(M)}]^T. \nonumber
\end{gather}

\section{Filtering and Smoothing Noisy Android Raw GNSS Measurements}
\subsection{PVT Solution Based on Moving Horizon Estimation}
MHE estimates the current system state from a moving window of data. Let $N+1$ denote the size of the moving window. Then, we need to determine the state at the $k^{th}$ epoch $\mathbf{X}_k$ with the measurements from the ${k-N}^{th}$ epoch to the $k^{th}$ epoch $\left[\mathbf{b}_{k-N},\mathbf{b}_{k-N+1},\cdots,\mathbf{b}_{k}\right]^T$. The state-transition matrix $\mathbf{A}_{k,k-1}$ defined by \eqref{eq:Ax}, \eqref{eq:At} and \eqref{eq:stateTransMatrix} is non-singular. Thus, according to \eqref{eq:stateUpdate}, the state at the ${k-1}^{th}$ epoch can be derived from the state at the $k^{th}$ epoch if the process noise $\mathbf{W}_{k-1}$ is ignored, i.e.,
\begin{equation}\label{eq:inverseStateUpdate}
    {\mathbf{X}}_{k-1}=\mathbf{A}_{k,k-1}^{-1}{\mathbf{X}}_{k}.
\end{equation}

If the measurement equation at the $k^{th}$ epoch is linearized at its approximation $\Tilde{\mathbf{X}}_k$, the measurement equation at the ${k-1^{th}}$ epoch can be linearized at $\Tilde{\mathbf{X}}_{k-1}$ that can be derived as follows:
\begin{equation}\label{eq:inverseApproxStateUpdate}
    \Tilde{\mathbf{X}}_{k-1}=\mathbf{A}_{k,k-1}^{-1}\Tilde{\mathbf{X}}_{k}.
\end{equation}

By recursively substituting \eqref{eq:inverseStateUpdate} and \eqref{eq:inverseApproxStateUpdate} into \eqref{eq:measurementEqn} at all $N+1$ epochs and ignoring the measurement noise, we can get the following measurement equation system:
\begin{equation}\label{eq:compactBatchMeasEq}
   \mathbf{Y}_{k,N} = \mathbf{M}_N(\mathbf{X}_{k}-\tilde{\mathbf{X}}_{k}) 
\end{equation}
where
\begin{gather}
    \mathbf{Y}_{k,N} = [\mathbf{b}_k^T,\mathbf{b}_{k-1}^T,\cdots,\mathbf{b}_{k-N}^T]^T \nonumber
\end{gather}
\begin{gather}
    \mathbf{M}_N = [(\mathbf{C}_{k})^T, (\mathbf{C}_{k-1}\mathbf{A}_{k,k-1}^{-1})^T,\cdots, \nonumber
    \\
    (\mathbf{C}_{k-N}\mathbf{A}_{k-N+1,k-N}^{-1}\cdots\mathbf{A}_{k-1,k-2}^{-1}\mathbf{A}_{k,k-1}^{-1})^T]^T. \nonumber
\end{gather}

We can solve \eqref{eq:compactBatchMeasEq} using the LS method and get
\begin{equation} \label{eq:batchLsSolution}
    \hat{\mathbf{X}}_{k} = \mathbf{M}_N^+\mathbf{Y}_{k,N}+\tilde{\mathbf{X}}_{k}.
\end{equation}

The approximate state $\tilde{\mathbf{X}}_{k}$ will be updated using the estimated state $\hat{\mathbf{X}}_{k}$ for the iterative computation from \eqref{eq:inverseStateUpdate} to \eqref{eq:batchLsSolution} until the convergence of $\hat{\mathbf{X}}_{k}$. Like the WLS algorithm, we can also weight the measurement equations \eqref{eq:compactBatchMeasEq}.

\begin{figure}[!t]
    \centering
    \includegraphics[width=0.48\textwidth]{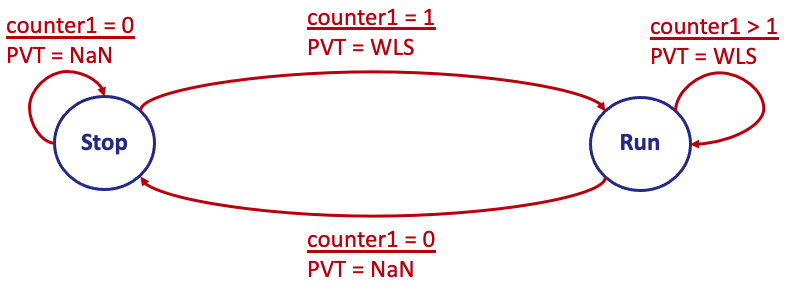}
    \caption{Finite state machine for WLS}
    \label{fig:fsmWls}
\end{figure}
\begin{figure}[!t]
    \centering
    \includegraphics[width=0.48\textwidth]{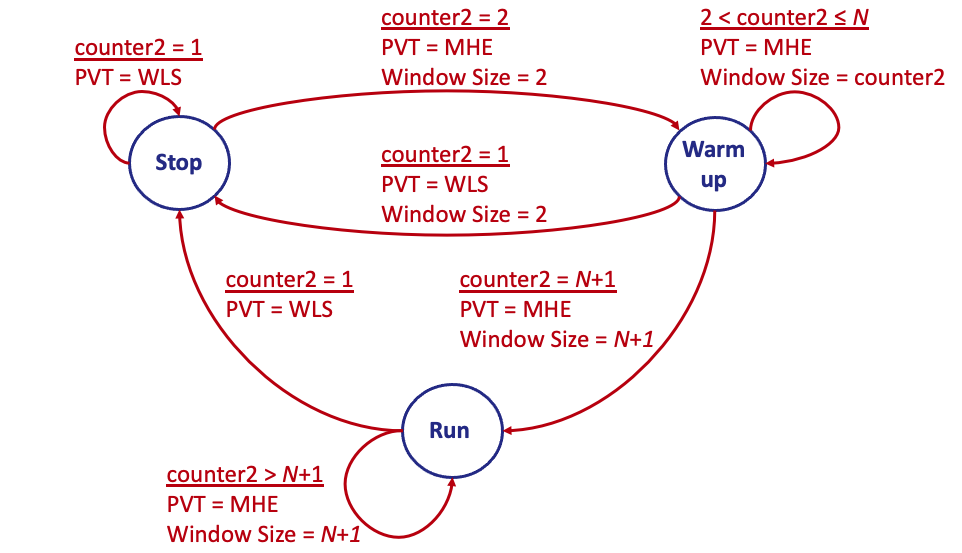}
    \caption{Finite state machine for MHE}
    \label{fig:fsmMhe}
\end{figure}

\subsection{PVT Solution Based on Extended Kalman Filter}
\subsubsection{General Algorithm}
EKF recursively estimates the current state based on all the previous data, which is illustrated by the following equations: 
\begin{gather}
    \hat{\mathbf{X}}_{k}^- = \mathbf{A}_{k,k-1}\hat{\mathbf{X}}_{k-1} \nonumber
    \\
    \mathbf{P}_{k}^- = \mathbf{A}_{k,k-1}\mathbf{P}_{k-1}\mathbf{A}_{k,k-1}^T+\mathbf{Q}_{k-1} \nonumber
    \\
    \mathbf{K}_k = \mathbf{P}_{k}^-\mathbf{C}_k^T\left(\mathbf{C}_k\mathbf{P}_{k}^-\mathbf{C}_k^T+\mathbf{R}_k\right)^{-1} \nonumber
    \\
    \hat{\mathbf{X}}_{k} = \hat{\mathbf{X}}_{k}^-+\mathbf{K}_k\mathbf{b}_k \nonumber
    \\
    \mathbf{P}_k = \left(\mathbf{I}-\mathbf{K}_k\mathbf{C}_k\right)\mathbf{P}_k^- \nonumber
\end{gather}
where $\hat{\mathbf{X}}_{k-1}$ and $\hat{\mathbf{X}}_{k}$ represent the posterior state estimation. $\mathbf{P}_{k-1}$ and $\mathbf{P}_{k}$ are the covariance matrix of the posterior state estimation. $\mathbf{P}_{k}^-$ denotes the covariance matrix of the prior estimation $\hat{\mathbf{X}}_{k}^-$. $\mathbf{K}_k$ represents the Kalman gain. Note that the approximate state $\tilde{\mathbf{X}}_k$ is replaced by the prior state estimation $\hat{\mathbf{X}}_k^-$ to calculate $\mathbf{b}_k$. $\mathbf{R}_k$ is the covariance matrix of measurement noise.

\begin{figure}[!t]
    \centering
    \includegraphics[width=0.48\textwidth]{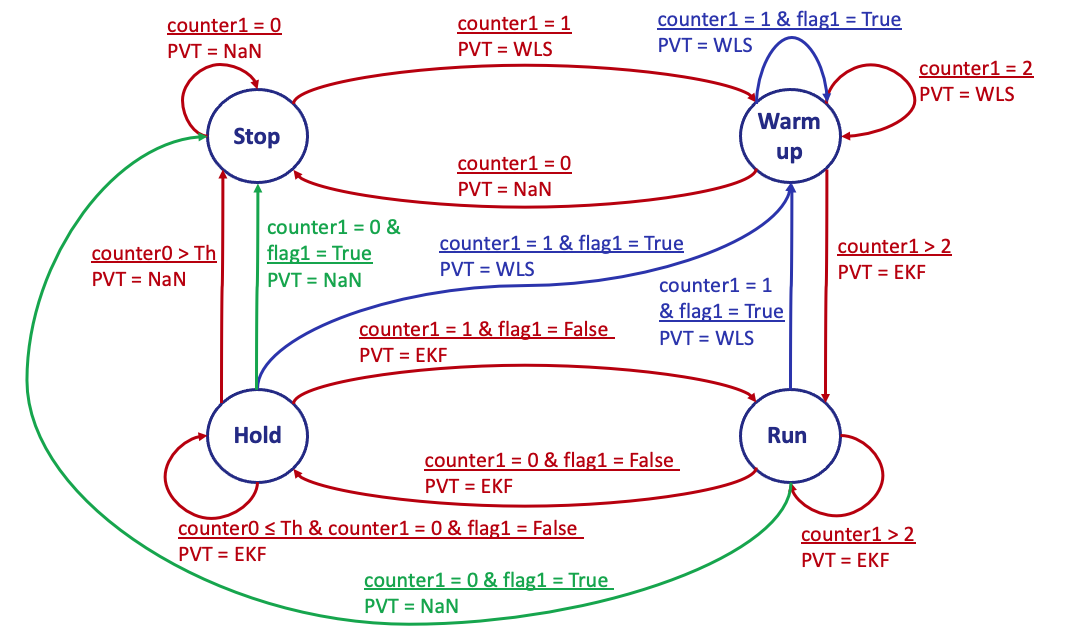}
    \caption{Finite state machine for EKF}
    \label{fig:fsmEkf}
\end{figure}
\begin{figure}[!t]
    \centering
    \includegraphics[width=0.48\textwidth]{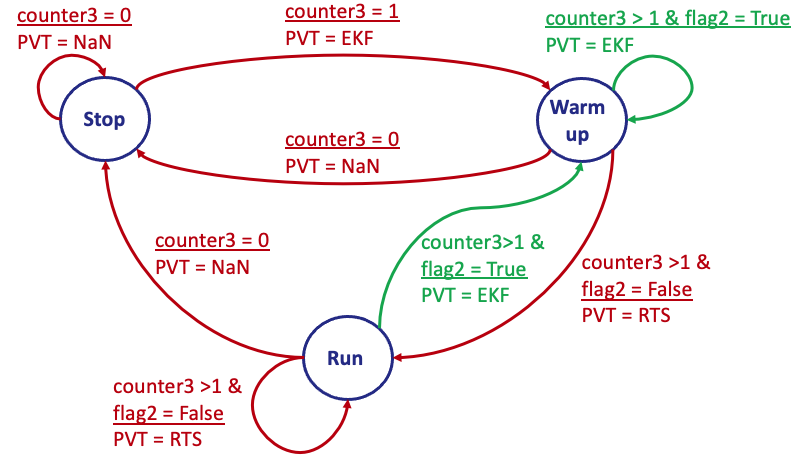}
    \caption{Finite state machine for RTS smoother}
    \label{fig:fsmRts}
\end{figure}

\subsubsection{Determination of Covariance Matrix $\mathbf{Q}_{k-1}$ and $\mathbf{R}_k$}
The computation of $\mathbf{Q}_{k-1}$ has been introduced in detail in Section \ref{sec:modeling}. Note that \eqref{eq:SxCompute}, \eqref{eq:StCompute}, and \eqref{eq:SfCompute} tell that the state estimation at the previous two steps is needed to determine the covariance matrix at the current epoch. The precedent states can be initially estimated with the WLS solutions and then gradually replaced with the EKF-based solutions.

Assume the pseudorange noise and pseudorange rate noise are unbiased and uncorrelated with each other. And the measurement noise is uncorrelated among various satellites. Then, $\mathbf{R}_k$ is calculated as follows:
\begin{gather}
    \mathbf{R}_k={\rm E}\left(\mathbf{E}_k\mathbf{E}_k^T\right)\nonumber
    \\
    =\begin{bmatrix}{\sigma_{\rho_k}^{(1)}}^2&0&0&0&0&\cdots&0&0\\
    0&{\sigma_{\dot{\rho}_k}^{(1)}}^2&0&0&0&\cdots&0&0\\
    \vdots&\vdots&\vdots&\vdots&\vdots&\vdots&\vdots&\vdots\\
    0&0&\cdots&0&0&0&{\sigma_{\rho_k}^{(M)}}^2&0\\
    0&0&\cdots&0&0&0&0&{\sigma_{\dot{\rho}_k}^{(M)}}^2\\
                \end{bmatrix} \nonumber
\end{gather}
where
\begin{IEEEeqnarray*}{rCl}
    &{\sigma_{\rho_k}^{(n)}}^2&={\rm E}\left({\varepsilon_k^{(n)}}^2\right)
    \\
    &&=ReceivedSvTimeUncertaintyNanos^2 
    \\
    &{\sigma_{\dot{\rho}_k}^{(n)}}^2&={\rm E}\left({\dot{\varepsilon}_k^{(n)^2}}\right)
    \\
    &=& PseudorangeRateUncertaintyMetersPerSecond^2.
\end{IEEEeqnarray*}

\begin{figure*}[!t]
\begin{subfigure}{0.24\textwidth}
\includegraphics[width=\linewidth]{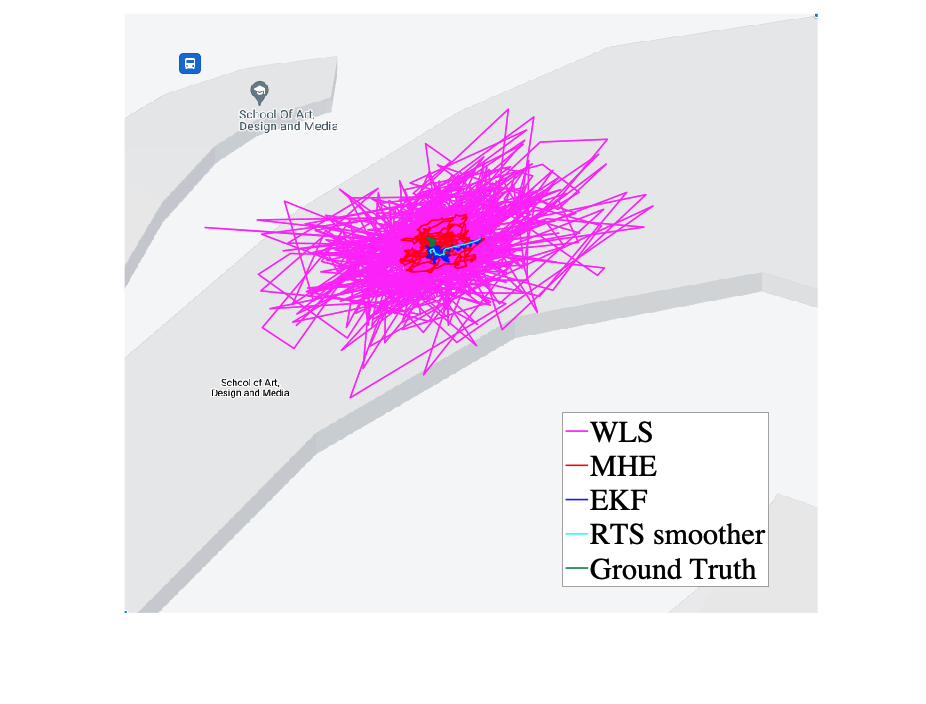} 
\caption{Traces}
\label{fig:Route1}
\end{subfigure}
\begin{subfigure}{0.24\textwidth}
\includegraphics[width=\linewidth]{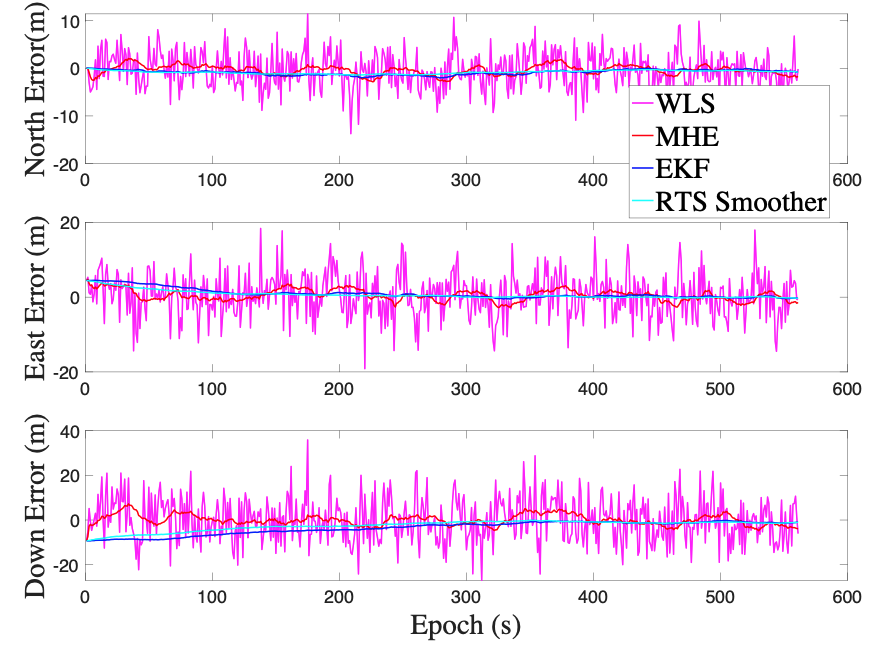}
\caption{Positioning Errors}
\label{fig:PosErr1}
\end{subfigure}
\begin{subfigure}{0.24\textwidth}
\includegraphics[width=\linewidth]{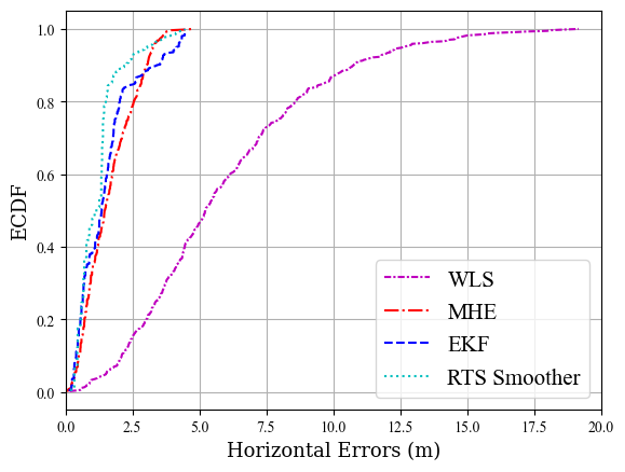}
\caption{Horizontal ECDF}
\label{fig:HECDF_1}
\end{subfigure}
\begin{subfigure}{0.24\textwidth}
\includegraphics[width=\linewidth]{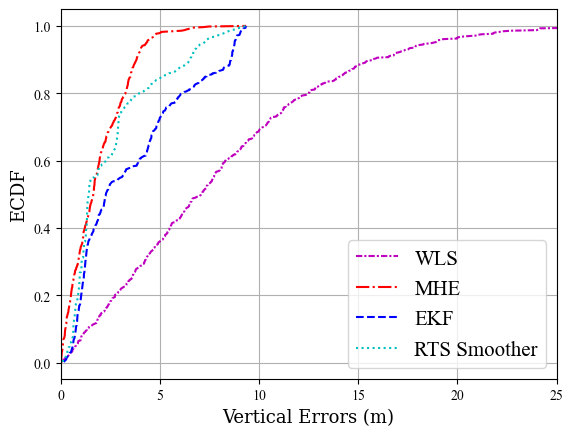}
\caption{Vertical ECDF}
\label{fig:VECDF_1}
\end{subfigure}
\caption{Evaluation of static scenario}
\label{fig:static}
\end{figure*}

\begin{figure*}[!t]
\begin{subfigure}{0.24\textwidth}
\includegraphics[width=\linewidth]{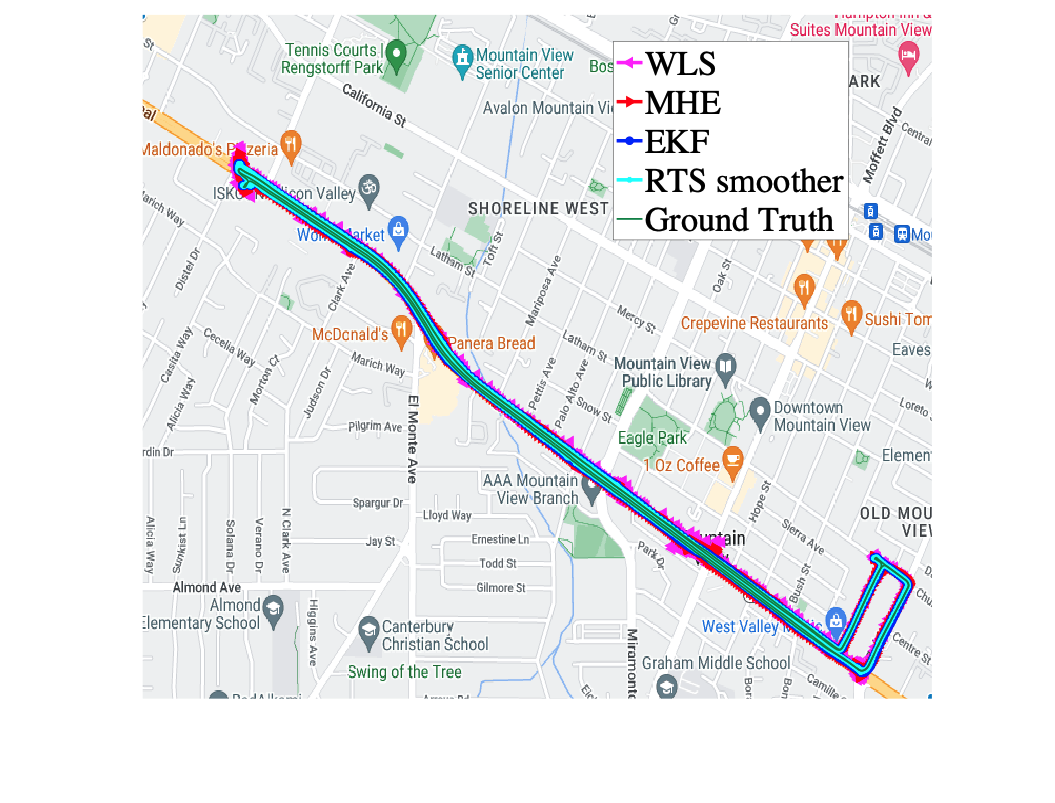} 
\caption{Traces}
\label{fig:Route2}
\end{subfigure}
\begin{subfigure}{0.24\textwidth}
\includegraphics[width=\linewidth]{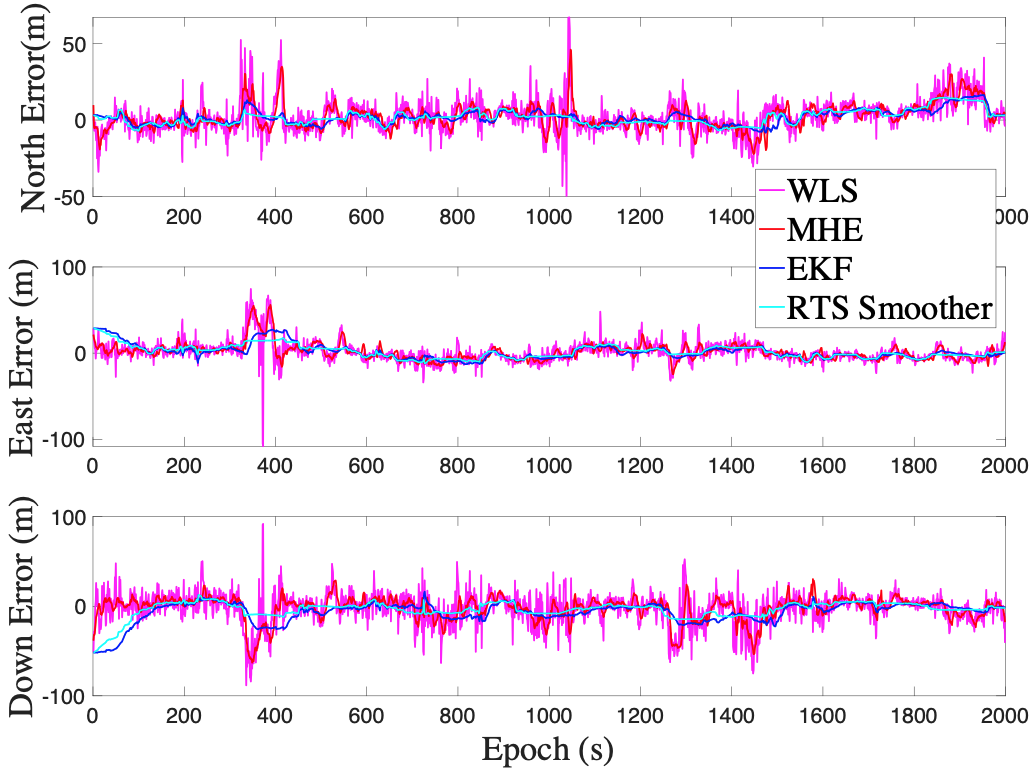}
\caption{Positioning Errors}
\label{fig:PosErr2}
\end{subfigure}
\begin{subfigure}{0.24\textwidth}
\includegraphics[width=\linewidth]{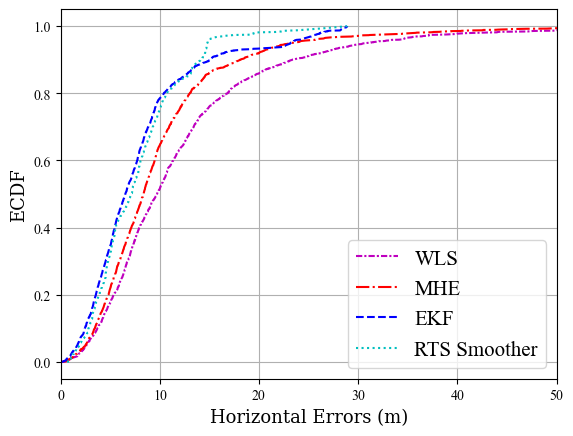}
\caption{Horizontal ECDF}
\label{fig:HECDF_2}
\end{subfigure}
\begin{subfigure}{0.24\textwidth}
\includegraphics[width=\linewidth]{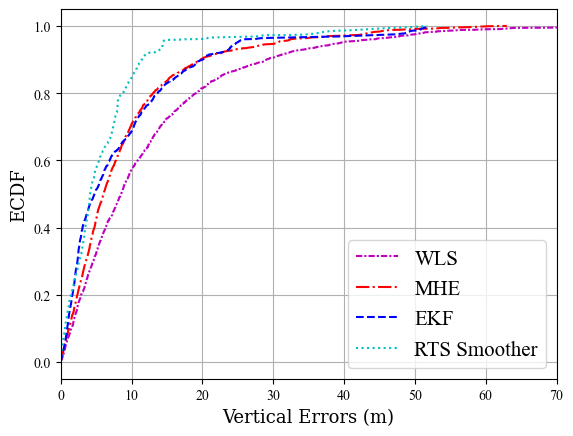}
\caption{Vertical ECDF}
\label{fig:VECDF_2}
\end{subfigure}
\caption{Evaluation of dynamic scenario}
\label{fig:dynamic}
\end{figure*}

\subsection{PVT Solution Based on Rauch-Tung-Striebel Smoother}
RTS smoother is a backward EKF, which starts from the last-epoch state estimated by EKF and smooths the state estimation backward. Thus, before using it, we should have obtained $\hat{\mathbf{X}}_k^-$, $\mathbf{P}_k^-$, $\hat{\mathbf{X}}_k$, and $\mathbf{P}_k$ using the forward EKF. Let $\hat{\mathbf{X}}_k^S$ and $\mathbf{P}_k^S$ represent the smoothed state estimation and the corresponding covariance matrix. The recursive formation of RTS smoother is shown below:
\begin{gather*}
    \hat{\mathbf{X}}_k^S = \hat{\mathbf{X}}_k+\mathbf{S}_k\left(\hat{\mathbf{X}}_{k+1}^S-\hat{\mathbf{X}}_{k+1}^-\right)
    \\
    \mathbf{P}_k^S = \mathbf{P}_k + \mathbf{S}_k\left(\mathbf{P}_{k+1}^S-\mathbf{P}_{k+1}^-\right)\mathbf{S}_k^T
\end{gather*}
where
\begin{gather*}
    \mathbf{S}_k = \mathbf{P}_k\mathbf{A}_{k,k-1}\left(\mathbf{P}_{k+1}^-\right)^{-1}.
\end{gather*}


\section{Practical Implementation for Discontinuous Data}
We detect three kinds of discontinuity in Android raw GNSS measurements. The first is localization failure due to fewer than four visible satellites, which we call satellite discontinuity. The second is the clock discontinuity of Android phones. That is, the time of two adjacent measurements is discontinuous. In this study, two adjacent data are considered discontinuous if the time difference exceeds 10 seconds. The last one is the pseudorange discontinuity. It means the pseudorange change between two consecutive epochs is larger than an expected value (such as 50 km), which may be caused by signal blocking, multipath, duty cycles, etc. We design finite state machines (FSM) to handle such discontinuity, as shown by Fig. \ref{fig:fsmWls}-Fig. \ref{fig:fsmRts}.

\subsection{Handling Data Discontinuity for the WLS Algorithm}
Only the satellite discontinuity influences the WLS algorithm, which is illustrated by its FSM containing two states, i.e., ``Stop" and ``Run," shown in Fig. \ref{fig:fsmWls}. ``counter1" stores the number of measurements which are collected from enough satellites. Once the satellite discontinuity happens, ``counter1" will be reset to zero, and the WLS algorithm will stop.

\subsection{Handling Data Discontinuity for MHE}
The satellite discontinuity rarely happens in MHE because MHE combines a window of data that generally guarantees enough visible satellites for PVT computation. Thus, we only consider the clock discontinuity and the pseudorange discontinuity for it. The combination of two measurements with these discontinuities will lead to large estimation errors. As shown in Fig. \ref{fig:fsmMhe}, we use ``counter2" to store how many measurements are continuous in terms of both time and pseudoranges. The ``Warm up" state means that MHE will run with data fewer than its preset window size. If any of the two kinds of discontinuity is detected, ``counter2" will be set back to one, and MHE will be replaced with WLS. 

\subsection{Handling Data Discontinuity for EKF}
All three kinds of discontinuity will affect EKF. The FSM of EKF is shown in Fig. \ref{fig:fsmEkf}, which involves four states, i.e., ``Stop," ``Warm up," ``Run," and ``Hold." ``flag1" indicates whether the clock discontinuity or pseudorange discontinuity takes place at the moment. A non-zero ``counter1" will be set to 1 once ``flag1" is true, i.e., the current measurement is discontinuous in clock or pseudoranges but contains enough visible satellites for WLS-based state estimation. ``counter0" counts how many measurements with satellite discontinuity have been accumulated. If satellite discontinuity happens, EKF will run in the ``Hold" state and infer the phone's state without any adjustment until counter0 is larger than ``Th". ``Th" represents the number threshold of consecutive satellite-discontinuity data and is set to 10.

\subsection{Handling Data Discontinuity for RTS Smoother}
RTS smoother estimates the current state based on the current and subsequent states given by EKF. Therefore, we only need to check whether the current EKF-based state estimation is empty and whether the current and the next EKF-based state estimations are continuous. As shown in Fig. \ref{fig:fsmRts}, ``counter3" counts how many consecutive non-empty states have been given by EKF backward to the current epoch. If the current EKF-based state estimation is empty, ``counter3" will be set back to 0. ``flag2" indicates whether the clock discontinuity or pseudorange discontinuity takes place at the next epoch. 

\section{Experiments}
We evaluate the performance of the aforementioned positioning algorithms in static and dynamic scenarios. The static data were collected by HUAWEI Mate 10 Pro on the roof of the School of Art, Design and Media at Nanyang Technological University, with ground truth collected by a u-blox receiver. For dynamic scenes, we use Google public datasets collected in Mountain View with Pixel 4 with ground truth provided by the NovAtel SPAN system \cite{fu2020android}. The positioning traces and errors are illustrated in Fig. \ref{fig:static} and \ref{fig:dynamic}. We score each method using the mean of the $50^{th}$ and $95^{th}$ percentile of horizontal errors computed by Vincenty's formulae, which is the evaluation metric in the Google Smartphone Decimeter Challenge (GSDC) and summarized in Table \ref{tab1}.

\begin{table}[htbp]
\caption{Horizontal positioning scores}
\begin{center}
\begin{tabular}{|c|c|c|}
\hline
\multirow{2}*{\textbf{Methods}}&\multicolumn{2}{|c|}{\textbf{Horizontal Score (meter)}}\\
\cline{2-3}
&\textbf{Static Scenario}&\textbf{Dynamic Scenario}\\
\cline{1-3} 
\textbf{WLS} & 8.9280&20.4651\\
\cline{1-3}
\textbf{MHE} & 2.3671&15.9024\\
\cline{1-3}
\textbf{EKF} & 2.7485&14.8676\\
\cline{1-3}
\textbf{RTS Smoother} & \textbf{2.1051}&\textbf{10.9495}\\
\cline{1-3}
\hline
\end{tabular}
\label{tab1}
\end{center}
\end{table}

As shown in Fig. \ref{fig:Route1}, Fig. \ref{fig:PosErr1}, Fig. \ref{fig:Route2}, and Fig. \ref{fig:PosErr2}, MHE, EKF, and RTS smoother significantly mitigate noise and obtain much smoother positioning results compared with the baseline WLS algorithm. Fig. \ref{fig:HECDF_1}, Fig. \ref{fig:VECDF_1}, Fig. \ref{fig:HECDF_2}, and Fig. \ref{fig:VECDF_2} display the empirical cumulative distribution function (ECDF) of horizontal and vertical positioning errors, demonstrating that MHE, EKF, and RTS smoother improve the positioning performance substantially. As indicated in Table \ref{tab1}, RTS smoother achieves the best horizontal score and reduces the horizontal localization error by $76.4\%$ and $46.5\%$ in static and dynamic scenes, respectively, compared to the WLS algorithm.

\section{Conclusion}
In this work, we detail how to compute locations using Android raw GNSS measurements and implement MHE, EKF, and RTS smoother to handle the large noise present in them. Besides, we devise dedicated finite-state machines for these localization algorithms to address the discontinuity in Android data. The experiment results show that the filtering or smoothing methods can significantly relieve the adverse impact of noise on localization results for Android smartphones. 

However, the positioning results in the dynamic scene indicate that these methods cannot eliminate the positioning bias (around 10 meters), which is the problem we need to solve in the next step to further improve the pseudorange localization performance of Android smartphones.

\bibliographystyle{unsrt}
\bibliography{main}

\end{document}